\documentstyle[aps,preprint]{revtex}
\begin{document}
\draft

\title{Stabilizing chaotic vortex trajectories: an example of
high-dimensional control}
\author{\'A. P\'entek$^1$,  J. B. Kadtke$^1$, and Z. Toroczkai$^{2,3}$}
\address{$^1$Institute for Pure and Applied Physical Sciences, 
University of California at San Diego, \\
9500 Gilman Dr., La Jolla, CA 92093-0360 \\ 
$^2$Department of Physics, Virginia Polytechnic Institute and
State University, \\
Blacksburg, VA 24061-0435\\
$^3$Institute for Theoretical Physics, E\"otv\"os University,\\
Puskin utca 5-7, H-1088 Budapest, Hungary}
\date{October 7, 1996}

\maketitle

\begin{abstract}
A chaos control algorithm is developed
to actively 
stabilize  unstable periodic orbits of higher-dimensional systems.
The method assumes knowledge of the model equations and 
a small number of experimentally accessible parameters. General
conditions for controllability are discussed. The algorithm is 
applied to the Hamiltonian problem of point vortices inside a circular
cylinder with applications to an experimental plasma system.
\end{abstract}

\newpage

\section{Introduction}

Hamiltonian models of 3-D collinear vortex filaments or 2-D point 
vortices inside closed geometries have for many years attracted 
interest due to its central role in understanding
the evolution of vorticity in real flows
\cite{Nov,Aref1,Aref,Kadtke,Franese,Pentek}.
2-D point vortex dynamics is also relevant in a series of real
experimental applications such as vortices in superfluid helium
\cite{helium}, superconductors,
and dynamics of non-neutral plasma filaments \cite{plasma}.
It has also been shown recently that point-vortex models can
provide a good formulation for developing control algorithms
in realistic flows \cite{KPP}, since they are able to 
capture in certain limits most of the qualitative features
of the vorticity dynamics.
These results indicate that control algorithms based
on these low-dimensional point vortex models can be successfully
implemented for certain geometries even in fully-viscid Navier-Stokes
equations \cite{PKP}.

In this Letter we will likewise attempt to develop a low-dimensional
model for a fluid control algorithm, aimed at modifying a realistic
flow of vorticity inside a bounded fluid domain. The model will be
based  on Hamiltonian point vortex dynamics, and make use of unstable periodic
trajectories in the natural flow dynamics to modify the flow
via the Ott, Grebogi, Yorke (OGY) \cite{OGY} scheme. The principal
result here is that we develop a general 
control algorithm which can be utilized
in higher-dimensional systems (phase space $D > 3$), such as
multi-vortex systems. We demonstrate the method's effectiveness
using numerical examples that model an experimental system of
confined non-neutral plasma.

To develop the control scheme, we first briefly review the
formulation of point vortex dynamics.
Let $z = x+i y$ denote the complex coordinate of a single vortex.
The effect of the boundary of the circular domain is
simply accounted for by image charges at $R^2/z^*$.
The Hamiltonian of a system of $N$ identical
point vortices each of circulation $\Gamma$ inside a circle of
radius $R$ is then given by \cite{Aref}
\begin{equation}
H(z_1,\ldots,z_N) = - \frac{\Gamma^2}{4 \pi} \left[ \;
\sum_{k \ne j}^N \; \ln{
\frac{\mid \! z_k - z_j \!\mid}
{\mid \! R^2-z_k z_j^* \! \mid}}\; - \;
\sum_k^N \; \ln{(R^2-\mid 
\! z_k \!\mid^2)} \; \right], \label{ham}
\end{equation}
Each vortex moves by being passively advected in the velocity field
created by the other vortices and the image vortices, with
the dynamical equations being obtained from the canonical 
Hamilton relations:
\begin{equation}
\Gamma \; \dot{z}_k^* = i\; \frac{\partial H}{\partial z_k},
 \;\;\;\;\;k= 1, \ldots, N,
\end{equation}
that is
\begin{equation}
\dot{z}_k^* = - \frac{i\Gamma}{4\pi} \left[ \;
\sum_{j\ne k}^N \frac{1}{z_k - z_j} \; + \;
\sum_{j}^N \frac{z_j^*}{R^2-z_j^* z_k}\; 
\right],
 \;\;\;\;\;k= 1, \ldots, N. \label{dyn}
\end{equation}
Here, the first sum includes the velocity field created by the
other vortices at the position of the $k$-th vortex 
(infinite self-interaction
excluded) while the second sum gives the velocity field
of the image vortices. The Hamilton relations (2) imply that
each vortex contributes one degree-of-freedom and
therefore two dimensions to the phase space. In addition, 
since the Hamiltonian does not contain  time explicitly,
the ``energy'' of the vortex system 
$E=H(z_1, \ldots,z_N)$ is a constant of the motion.
There exists another constant of motion 
for a circular domain resulting from the invariance of $H$
under rotation, i.e. the ``angular momentum'':
\begin{equation}
L = \Gamma \; \sum_k^N \mid\! z_k \!\mid^2 . \label{ang}
\end{equation}
Because of these two integrals of the motion, the dynamics 
of the $N$-vortex system is restricted to a
$2N-2$ dimensional manifold in the $2N$ dimensional phase space. 

The Hamilton relations for point vortices imply the unusual situation
that the phase space is simply a re-scaled version of the
configuration space. Hence, integrability conditions which
normally rely on symmetries of the Hamiltonian via Noether's
theorem, can partly be deduced by looking for symmetries in the
physical domain \cite{Nov,Aref1,Aref}.
In this view, one has
global non-integrability (chaos) if
$N^* \ge 1$, where $N^* = $ (number of vortices) - (number of remaining
integrals of motion) \cite{Nov}. For the circular domain discussed here,
$N^* \ge 1$ for $N \ge 3$. In the following we will
restrict our discussion to the three-vortex case, which is the
smallest number of vortices inside a cylinder that leads to
chaotic dynamics.

Our central aim here is to achieve control and avoid chaos by
stabilizing simple unstable periodic orbits of the
three-vortex system, thereby taking advantage of the
natural dynamics of the flow. Although there are a large number of low-order
periodic orbits of the generic vortex dynamics, we will restrict ourselves
without loss of generality
to those associated with symmetric configurations of vortices, with the
center-of-vorticity in the origin. Examples of such
configurations are shown in Fig.~1 for $N=3$. 
We can simplify the discussion
by rendering the equations-of-motion dimensionless 
with the substitution:
\begin{equation}
z \longrightarrow \frac{z}{R}, \;\;\;\;\; \mbox{and}
\;\;\;\;\;
t \longrightarrow t \;\frac{2\pi}{\Gamma R^2}.
\end{equation}
Performing a further transformation into a 
reference frame co-rotating with the vortex system
with uniform angular velocity $\omega$,
again without loss of generality, Eq.~(\ref{dyn}) will read as: 
\begin{equation}
\dot{z}_k^* = - \frac{i}{2} \left[ \;
\sum_{j\ne k}^3 \frac{1}{z_k - z_j} \; + \;
\sum_{j}^3 \frac{z_j^*}{1-z_j^* z_k}\; - \; 2
\omega z_k^* \;
\right],
 \;\;\;\;\;k= 1, 2, 3.
\end{equation}
The main advantage of this co-rotating frame is that some of the
periodic orbits
in the 6-D phase space reduce to a single fixed point (with
proper choice of $\omega$). 

The stability of such symmetric
configurations has been first analyzed by Havelock \cite{Hav}.
In spite of the global non-integrability of the system, the 
symmetric orbits are stable for some 
parameter regimes, i.e. they are elliptic
quasiperiodic orbits embedded in the four-dimensional 
manifold determined by the
constants of motion (\ref{ham}) and (\ref{ang}). 
These orbits correspond to states
where the vortices are away from the boundary of the cylinder,
hence
the effect of the ``image'' vortices is  a weak
perturbation, of the integrable three-vortex dynamics in free space. 
These regions are separated from the chaotic areas
by KAM tori containing orbits of marginal stability.  
Many of these stable configurations have been found experimentally
in the dynamics of magnetically confined non-neutral plasma columns
\cite{plasma}, 
where the dynamics is very close to
those of point vortices \cite{Levy}.

\section{Formulation of a high-dimensional control algorithm}

Although many examples of chaos control algorithms have been developed
in recent years, these have  been restricted to  a relatively
low-dimensional phase space. Here, we develop
a control algorithm for higher-dimensional systems which assumes
explicit knowledge of the dynamical equations and therefore
the location, eigenvalues and eigenvectors of the fixed
point or periodic orbit to be stabilized. This method can also certainly
be reformulated in the language of 
experimentally measured time series (as in previous works 
\cite{SO,Ding,Petrov} ), however
this is not our immediate aim here.

For the particular application we discuss 
in this Letter, we aim to modify
the dynamics of the fully-viscid continuous fluid
using a control model based on the point vortex dynamics. Hence,
the model governing equations are known, and the high-dimensional
control is straightforward to calculate. In the case of an experimental
system, the method can be implemented in an identical fashion, using the
Jacobian and the perturbation matrix obtained from experimental
time-series \cite{SGOY}.

To begin, let ${\dot{\bf r}} = {\bf A}({\bf r},{\bf \Phi})$ represent
the $n=2N$-dimensional unperturbed dynamical system of Eq.~(6). 
Here, in addition to Eq. (\ref{dyn}) we have introduced a 
set of $m$ experimentally
accessible system parameters, represented by the $m$-dimensional vector 
$\bf \Phi $. 
We assume the number  of such
parameters $m$ is usually much smaller  then the 
dimensionality of the phase space $n$, typically one.
  
Suppose that ${\bf r}_0$ is 
an unstable fixed point in the $n$-dimensional
phase space, i.e.,
$\dot{\bf r}_0 = {\bf A}({\bf r_0},{\bf \Phi_0}) = 0$,
where ${\bf \Phi_0}$ is a fixed set of parameters of interest.
Allowing small perturbations
$\delta {\bf \Phi}(t)$ in the parameters,
then in a neighborhood of this fixed point 
${\bf r} = {\bf r}_0 + \delta{\bf r}$ the dynamics is described by
the linearized equations
\begin{equation}
\delta\dot{\bf r}(t) = {\bf J}({\bf r}_0, {\bf \Phi}_0) \; 
\delta{\bf r}(t)
+ {\bf G}({\bf r}_0, {\bf \Phi}_0) \;\delta {\bf \Phi}(t) \label{dev}
\end{equation}
where ${\bf J} = \partial {\bf A}/\partial {\bf r}$
is the standard Jacobian and the $n\times m$ matrix 
${\bf G} = \partial {\bf A}/\partial {\bf \Phi}$
describes the effect of small parameter perturbations on the system
(i.e. the perturbation matrix).

A stability analysis can be performed around the fixed point
${\bf r}_0$, by studying the properties of the Jacobian, and 
typically can reveal a multitude of topologies. These topologies are 
defined by the set of eigenvalues of $\bf J$, which can be
real, complex or zero.  
The instability of the fixed point implies that there is
at least one eigenvalue $\lambda$ with $Re(\lambda) > 0$.
Since the system is Hamiltonian 
($Tr \; {\bf J} = 0$)
there is also at least one eigenvalue with
negative real part. Although there is no guarantee of purely real eigenvalues
and eigenvectors,
with a proper transformation one can eliminate the imaginary part
for at least one of the stable eigenvalues (and the corresponding 
eigenvector) without changing the stability.
In the following, we will suppose without loss
of generality that there is at least one purely real negative eigenvalue
$\lambda_s < 0$ with a corresponding normalized, stable real 
eigenvector ${\bf e}_s$. 
 
To achieve control of the vortex trajectories using small perturbations
$\delta {\bf \Phi}$, we impose the condition that after a short time
$\Delta t$ the trajectory 
has approached the fixed point, i.e. 
$\mid\delta{\bf r}(t+\Delta t)\mid<\mid\delta{\bf r}(t)\mid$ .
This can generally be accomplished in many different ways. 
One possible criteria  is 
suggested by the low-dimensional chaos control 
of OGY \cite{OGY},
where the new point $\delta{\bf r}(t+\Delta t)$ 
is driven onto the  stable manifold of the fixed point.
Other possible choices are the self-locating (geometric control) 
method originally developed
for low-dimensional chaos  \cite{geom1,geom2}
or a method using a Newton algorithm developed for 
higher-dimensional systems \cite{Newton}.
Here, we adopt the approach of OGY , i.e. we require
that the trajectory lies on the stable manifold of the fixed point
after a time $\Delta t$ \cite{KPP}:
\begin{equation}
\delta{\bf r}(t+\Delta t) = 
\alpha \mid\! \delta{\bf r}(t) \!\mid {\bf e}_s \label{cri}
\end{equation}
where alpha is a small real number ($0 < \alpha <1$).
Intuitively, this implies that after time $\Delta t$ the trajectory
lies on the stable manifold and simultaneously 
the distance from the fixed point
has been decreased by $\alpha$. Using a discretization
of (\ref{dev}),
this equation can be written explicitly as
\begin{equation}
\alpha \mid\! \delta{\bf r}(t) \!\mid {\bf e}_s \simeq
[{\bf 1} + {\bf J} \Delta t] \; \delta{\bf r}(t)
+ {\bf G}\;\delta{\bf \Phi}(t) \Delta t,  \label{con}
\end{equation}
where for simplicity the notations ${\bf J}\equiv
{\bf J}({\bf r}_0, {\bf \Phi}_0)$ and ${\bf G}\equiv
{\bf G}({\bf r}_0, {\bf \Phi}_0)$ have been introduced.
This represents a system of $n$ linear equations with $m$ unknown
perturbation parameters $\delta {\bf \Phi}$
that typically has no solution
when $m < n$. This means that it is not possible to control
such a system with one or a few control parameters in a one-step
process, as described by Eq. (\ref{con}).

Here, however we develop an alternative way
to overcome this difficulty.
Let us introduce the following notation for the RHS of Eq.~(\ref{con}):
\begin{equation}
{\bf F}({\bf r},{\bf \Phi}) \equiv 
[{\bf 1} + {\bf J} \Delta t] \;\delta{\bf r}(t)
+ {\bf G}\; \delta {\bf \Phi}(t) \Delta t.
\end{equation}
Then, instead of Eq.~(\ref{cri}) one can impose a weaker condition: namely,
that after $p$ steps the trajectory lies on the stable
manifold, i.e.
\begin{equation}
\delta{\bf r}(t+p\Delta t) = 
\alpha \mid\! \delta{\bf r}(t) \!\mid {\bf e}_s \label{wek}
\end{equation}
or
\begin{equation}
\alpha \mid\! \delta{\bf r}(t) \!\mid{\bf e}_s = {\bf F}( {\bf F}
(\ldots ({\bf F}({\bf r},\delta{\bf \Phi}^{(1)}), \ldots ),
\delta{\bf \Phi}^{(p-1)}), \delta{\bf \Phi}^{(p)}) \label{iter}
\end{equation}
where the shorthand notation $\delta {\bf \Phi^{(k)}}$ has been introduced
for the perturbation at time $t+k\Delta t$, i.e. $ {\bf \Phi}(t+k\Delta t)$.
Eq~(\ref{iter}) then can be written explicitly as
\begin{equation}
{\bf M}^{p-1}  {\bf G} \; \delta{\bf \Phi}^{(1)} + {\bf M}^{p-2} {\bf G}\; 
\delta{\bf \Phi}^{(2)} + \ldots + {\bf G} \;\delta{\bf \Phi}^{(p)}
   = \frac{1}{\Delta t} \left(
\alpha \mid\! \delta{\bf r}(t) \!\mid {\bf e}_s - {\bf M}^p \delta{\bf r}(t)
\right)
\end{equation}
where ${\bf M} \equiv {\bf 1} + {\bf J} \Delta t$.
This provides us with $n$ linear equations and $\;\;m p\;\;$ unknown
perturbations. The number $p$ is choosen as the smallest integer satisfying 
the condition $mp \ge n$. In the following, we will assume that there is a
$p$ such as $m p = n$. If $n$ is not divisible by $m$, there are
$m p -n$ perturbation parameters that can be freely set to zero.

We note that the size $\delta r^*$  of 
the linear neighborhood  of the fixed point
gives a natural limit for the allowed size of the perturbations,
and we assume that none of the perturbation
parameters $\delta \Phi_k$ can be larger 
than a preselected value $\delta\Phi^*$.
However, since  Eq.~(10) contains the yet unspecified
control time parameter $\Delta t$, the perturbations scale with
this parameter as $1/\Delta t$. This means that we should impose
a limit on the product $\delta \Phi_k \Delta t$ rather than
on $ \delta\Phi_k$ alone. 
The upper bound of this product should naturally be set by the system's 
physical limitations. 
For example, in the 3-vortex problem
$\Phi_k \Delta t$ is a length that is naturally limited
by the typical size of the linearized region $\delta r^*$, i.e.
$\delta \Phi^* \Delta t = \delta r^*$.

It is possible to derive general 
conditions for the controllability of the system.
To have a  solution for the $p$-step control process (13), 
the square matrix ${\bf C}$  
formed
from the $p$ matrices of size  $n\times m$, i.e. 
\begin{equation}
{\bf C} \equiv
\left\{ {\bf M}^{p-1} {\bf G} ; {\bf M}^{p-2} {\bf G} ; \ldots ; 
{\bf G} \right\}
\end{equation}
has to have nonvanishing determinant.
In two-dimensions, the system is always 
controllable provided the vector ${\bf G}$
is not collinear with ${\bf e}_s$. Intuitively, this means that
small perturbations of the system do not move the fixed point exclusively 
along the 
stable direction. In higher-dimensions, such a condition is not 
strong enough;
the perturbation may access only a very limited subspace, or
may have a certain symmetry that leads 
to un-controllability, even if the two-dimensional analog condition is
satisfied. Instead  
\begin{equation}
\mbox{det}\;{\bf C} \neq 0 \label{cond}
\end{equation} 
incorporates all these additional
conditions, and can be viewed as the basic criteria of controllability.
This relation is also a useful  tool to select appropriate 
perturbations for each particular
control problem, and to find the minimum number of perturbations
that can successfully control the system. 
The control parameter selection is probably the most important step
in any such problem,
especially when the Jacobian is highly-symmetric as in the case of 
symmetric periodic orbits of identical vortices. 
For such orbits it is straightforward to show that
most of the symmetric perturbations lead to un-controllability.
An example of this is the simple perturbation
of ``squeezing'' the boundary, i.e.
making it slightly elliptic with a small eccentricity.
Here, the equations of
motion can be simply deduced by using a simple conformal mapping
$f(z) = z+\epsilon z^3$
that maps the unit circle into an ellipse, with semi-axis $1+\epsilon$
and $1-\epsilon$, respectively (for $\epsilon \ll 1$).
A careful analysis of the collinear configurations of Fig.~1(a) shows
that the determinant of the corresponding controllability matrix $\bf C$
vanishes, due to the symmetry of the Jacobian $\bf J$ and the perturbation
matrix $\bf G$ with respect to the non-central vortices.
This means that it is not possible to control this particular configuration
by using  cylinder squeezing. 
In fact such behaviour is expected
to be quite general in systems of interacting identical  subsytems
that respond in the very same way to external perturbations.

\section{Numerical experiments}

To illustrate the above control algorithm, we actively stabilized the two
simplest unstable periodic orbits of the three vortex system inside a
cylinder, i.e. the symmetric states shown in Fig.~1(a,b).
For the collinear state of Fig.~1(a), the angular velocity of the
configuration is $\omega = (3+a^4)/(1-a^4)/(4a^2)$, while for the
triangular state of Fig.~1(b), it is $\omega = (1+2a^6)/(1-a^6)/(2a^2)$.
Here $a$ is the radial
distance of the non-central vortices from the center  of
the cylinder.  
To simplify the equations of motion, the dynamics is viewed in
a reference frame co-rotating with the vortices with this constant 
angular velocity
$\omega$. In this frame, the periodic orbits become  single
fixed points. Analytic calculation of the Jacobian for the
collinear state at $a=0.5$ reveals
the existence of four real eigenvalues, $\lambda_{1,2} = \pm 4.7565$
and $\lambda_{3,4} = 0$, and two purely complex ones,
$\lambda_{5,6} = 1.6027  i$. 
The triangular state at $a=0.7$ leads to two purely real eigenvalues
$\lambda_{1,2} = \pm 0.0001165$ and four complex ones,
$\lambda_{3,4,5,6} = \pm 0.6168 \pm 0.1670 i$.
The stable eigendirection ${\bf e}_s$
is then obtained in each case 
as the eigenvector associated with the single negative
real eigenvalue. One can observe that although the two topologies 
are completely different, the controller requires 
knowledge of only one of the possibly many stable directions. 

To control the vortex dynamics we require non-symmetric
and non-homogeneous perturbatios.
We have introduced simple 
perturbations applied only
at the cylinder surface: namely, a uniform distribution of sources and sinks
with variable strength $s = s_0 \cos{(k \theta)}$. 
Although such a perturbation may be difficult to implement in a real
fluid experiment, it is a real possibility in the plasma analog
of the vortex dynamics \cite{plasma}, 
where similar perturbations can be generated by external
electric fields. 
For $k=1, 2, \mbox{and}\; 3$, this leads to 
a uniform, a quadrupole, and a sextapole field, respectively,
as shown in Fig.~2(a-c).
Since all these fields are highly symmetric, none of them alone can 
effectively control the vortex dynamics.
However, their linear combinations, 
resulting in strongly asymmetric and inhomogenous fields,
proved to be a successful choice that 
satisfies the controllability conditions (\ref{cond}).
Here, we have introduceed three linearly independent external fields
as shown in Fig.~2(d-f). Using these, the equations
of motion can be written:
\begin{eqnarray}
\dot{z}_k^*  & = &  - \frac{i}{2} \left[ \;
\sum_{j\ne k}^3 \frac{1}{z_k - z_j} \; + \;
\sum_{j}^3 \frac{z_j^*}{1-z_j^* z_k}\; - \; 2
\omega z_k^* \;
\right] \nonumber \\
 & & +
\Phi_1 \; (1+z_k+z_k^2)+
\Phi_2 \;(-1+z_k-z_k^2)+
\Phi_3 \; (1+z_k-z_k^2)
 \;\;\;\;\;k= 1, 2, 3.
\end{eqnarray}

We previously showed that
there are two constants of motion in the unperturbed system,
the energy and the angular momentum. Due to the presence of 
the time-dependent 
perturbations that do not preserve these first integrals, however, 
the effective dynamics is no longer restricted to a 4-D manifold, but 
rather can occupy in the full 6-D phase space. 
Note that
if the dynamics initially lies in the same energy and angular momentum
manifold that contains our fixed point, ergodicity ensures that the
trajectory sooner or later reaches a close neighborhood
of the desired fixed point, and then the  control algorithm is switched 
on. 

If this is not the case, then in general 
this procedure can be supplemented with a targeting algorithm
that first drives the system close to the proper energy and 
angular momentum values, then takes advantage of the infinite
number of unstable periodic orbits in that manifold to
reach the neighborhood of the desired fixed point in a short time
\cite{target}. The development of such a targeting algorithm
in higher dimensions is clearly an important step
that deserves further study. 

Since this is not the primary emphasis of this Letter,  
in our numerical simulations we have started the system close to the
unstable fixed point. To demonstrate this
Fig.~3(a-b) shows the time evolution of a single coordinate
pair $(x_1, y_1)$ of the vortex system
in the collinear case, at first  without control (up to mark A). Then,
after the dynamics reaches  a small neighborhood of the 
desired fixed point (mark B), 
the controller is switched on, and the
trajectory is stabilized. Fig.~3(c-e) also shows the time evolution
of the required perturbations. Clearly this demonstrates that the algorithm
can effectively control the dynamics with only tiny perturbations
applied on the boundary. For better visualization, Fig.~4
shows a three-dimensional projection of the phase space trajectory
with and without control.

As previously mentioned, other control criteria can be used
instead of the OGY method. For example, one can consider the
high-dimensional control of Xu and Bishop, based on the
Newton root finding algorithm \cite{Newton}.
This method does not
assume knowledge of the stable direction, and it can be useful in cases
where all eigenvalues are complex (with at least two having non-vanishing real
part). Then, instead of condition (11) one can use the usual
Newton root finding algorithm
\begin{equation}
{\bf r}(t+p\Delta t) = {\bf r}(t) - {\bf J}^{-1}({\bf r}_0)  \;
{\bf A}({\bf r_0 + \delta r}). \label{Nw} 
\end{equation} 
However, this method fails when the Jacobian is not invertible.
(We note here that the transition from 
Eq.~(5) to Eq.~(6) in Ref. \cite{Newton}
is mathematically incorrect, i.e. the solution (6) does
not fulfill Eq. (5). Therefore the weaker
condition (\ref{Nw}) shown above 
should be used.) 

In a realistic Navier-Stokes simulation our algorithm
is effective on time scales significantly shorter than the viscous one
($Re^{1/2}$). This is attractive since this time scale
can be rather large in some
plasma systems, where the viscosity is small and there
are no boundary-layer effects due to the free-slip
conditions on the boundary. Even in these circumstances,
it should be noted that there
is an additional non-viscous effect that is not present in 
the Hamiltonian model, the vortex merger due to
the finite size of the vortices. Therefore effective
control can be reached only with concentrated vortices that
are far from each other, typically with 
$2\rho/d \gtrsim 1.8$, where $d$ is the distance between
vortices and $\rho$  the vortex radius computed according
to Ref.~\cite{plasma1}. Although this condition can be easily
fulfilled while the controller is working, during the targeting
algorithm or full chaotic dynamics the vortices may easily come close 
to each other. In this context, the principal usefulness of such
a control scheme will be to {\em prevent} the vortex merger on
short time scales from symmetric unstable initial conditions.  

In conclusion, we have demonstrated that the 
control algorithm derived in this Letter can effectively control
higher-dimensional systems. 
We have also derived a  controllability condition that does
not depend on the particular control method used, since it contains
only the Jacobian and the perturbation matrix as input parameters.
In this sense, the result is quite general, and can be used 
to decide the type and number of perturbations needed to 
control a higher-dimensional system.

For the particular numerical example presented 
here, i.e. three-vortex dynamics inside a cylinder, the
controller was designed in such a way that it could be implemented
in a magnetically confined plasma experiment. 
At present, we also plan to implement the control 
algorithm in
a realistic viscous fluid framework, and hence
an effective targeting algorithm has to be developed.
In addition, in the present paper the controller 
was formulated using exactly known
vortex coordinates.
However, the control scheme itself can be entierly reformulated
in a phase space reconstructed purely from wall signal (e.g. 
boundary pressure
or voltage) measurements,
thereby providing an algorithm which can be directly implemented in
an experimental framework.

After the acceptance of this paper for publication it was brought
to our attention a different formulation of the control algorithm that yields
to a similar controlability matrix \cite{Cels}.

\section*{Acknowledgements}

The authors wish to thank F. Driscoll, G. Pedrizzetti and T. T\'el 
for very useful discussions.
AP and JK wish to acknowledge partial support for this project by a grant
from the Office of Naval Research, No. N00014-96-1-0056, U.S.-Hungarian
Science and Technology Joint Fund under Project JFNo. 286 and 501,
and by the Hungarian Science Foundation under Grant Nos. OTKA
T17493, F17166. ZT wishes to acknowledge
R.K.P. Zia and B. Schmittmann for their support and
permanent encouragement. ZT has also been sponsored by the National
Science Foundation through the Division of Material Research.



\begin{figure}
\caption{Symmetric vortex configurations leading to
periodic orbits. Three-vortex configurations:
(a) collinear with one vortex in the center;
(b) equilateral triangle. This configuration leads to
stable dynamics for $a<0.567$, and unstable one for
$a>0.567$.  
Each configuration is rotating with an angular velocity 
$\omega$ that depends on the particular configuration. }
\end{figure}

\begin{figure}
\caption{Perturbation fields used to control the vortex dynamics.
Figs.~(a-c) show the velocity fields generated by a distribution
of sources and sinks on the boundary 
of strength $\cos{(k\theta)}$ for $k =1, 2, 3$,
respectively. Figs.~(d-f) displays the velocity fields resulting
from different linear combinations of the fields (a-c):
(d) $\dot{z}^* = \Phi_1\; (1+z+z^2)$, 
(e) $\dot{z}^* = \Phi_2\; (-1+z-z^2)$, and
(c) $\dot{z}^* = \Phi_3 \; (1+z-z^2)$. 
The strength of these fields $\Phi_k, k=1, 2, 3$
are computed by the control algorithm.}  
\end{figure}

\begin{figure}
\caption{(a) Vortex trajectories in the lab frame and 
the applied perturbations.
Figs.~(a-b) show the coordinates of one of the vortices,
with the collinear configuration started close to the fixed point.
Control is off until time $t=20$ (mark A). At this point the
controller is activated and control begins at 
mark B. Without a targeting algorithm,
the time between A and B can be very long. 
From instant B,
the controller applies the necessary perturbations at intermittent
times $\Delta t = 0.1$ to keep
the trajectory on the unstable periodic orbit. 
The parameter $\alpha$ in Eq.~(13) was 0.1 .
Figs.~(c-e) show the applied perturbations. 
After a short transient time the perturbations are
quite small, invisible on this scale.
Fig.~(e), inset,  also displays the perturbation $\Phi_3$
magnified 1000 times between $30< t< 40$, for visualization.} 
\end{figure}

\begin{figure}
\caption{
Three-dimensional projection 
($x_1$, $y_1$, $r_{12} = \mid \! z_2 - z_1\! \mid$) 
of the phase space trajectory
in the lab frame. 
(a) Uncontrolled trajectory of the collinear configuration.
(b) Trajectory after controller is switched on.
}
\end{figure}


\begin{references}

\bibitem{Nov} E. A. Novikov, and Yu. B. Sedov, Sov. Phys. - JETP {\bf 48},
440 (1978).  

\bibitem{Aref1} H. Aref, Phys. Fluids {\bf 22}, 393 (1979).

\bibitem{Aref} H. Aref, Ann. Rev. Fluid Mech. {\bf 15}, 345 (1983).

\bibitem{Kadtke} J. B. Kadtke, {\em PhD Thesis}, Brown University,
(1987).

\bibitem{Franese} L. Zannetti, and P. Franese, Euro. J. Mech. B/Fluids
{\bf 12}, 43 (1993).

\bibitem{Pentek} A. Pentek, T. Tel, and Z. Toroczkai, J. Phys. A {\bf 28},
2191 (1995); Fractals {\bf 3}, 33 (1995).

\bibitem{helium} P. H. Roberts, and  R. J. Donnelly, 
Ann. Rev. Fluid Mech. {\bf 6},
179 (1974).
 
\bibitem{plasma} K. S. Fine, A. C. Cass, W. G. Flynn, and
F. Driscoll, Phys. Rev. Lett. {\bf 75}, 3277 (1995).

\bibitem{KPP} J. B. Kadtke, A. Pentek, and G. Pedrizetti,
Phys. Lett. A {\bf 204}, 108 (1995). 

\bibitem{PKP} A. Pentek, J. B. Kadtke, and G. Pedrizzetti,
{\em Controlled capture of vortices in open viscous flows},
preprint (1996).

\bibitem{OGY} E. Ott, C. Grebogi, and J. A. Yorke, Phys. Rev. Lett.
{\bf 64}, 1196 (1990).

\bibitem{Hav} T. H. Havelock, Phyl. Mag. S. 7 {\bf 11}, 617 (1931).

\bibitem{Levy} R. H. Levy, Phys. Fluids {\bf 8}, 1288 (1965).

\bibitem{SO} P. So, and E. Ott, Phys. Rev. E {\bf 51}, 2955 (1995).

\bibitem{Ding} M. Ding, et al., Phys. Rev. E {\bf 53}, 4334 (1996).

\bibitem{Petrov} V. Petrov, E. Mihaliuk, S. K. Scott, and K. Showalter,
Phys Rev. E {\bf 51}, 3988 (1995).

\bibitem{SGOY} T. Shinbrot, C. Grebogi, E. Ott, and J. A. Yorke,
Nature {\bf 363}, 411 (1993).

\bibitem{geom1} Z. Toroczkai, Phys. Lett. A {\bf 190}  71 (1994).

\bibitem{geom2} B. Sass, and Z. Toroczkai, J. Phys. A {\bf 29}, 3545 (1996).

\bibitem{Newton} D. Xu, and S. R. Bishop, Phys. Lett. A {\bf 210},
273 (1996).

\bibitem{target} T. Shinbrot, et al., Phys. Rev. Lett. {\bf 68},
2863 (1992).

\bibitem{plasma1} K. S. Fine, C. F. Driscoll, J. H. Malmberg, 
and T. B. Mitchell,
Phys. Rev. Lett. {\bf 67}, 588 (1991).

\bibitem{Cels} F. J. Romeiras, C. Grebogi, E. Ott, and W. P. Dayawansa,
Physica D {\bf 58}, 165 (1992).
\end{references}
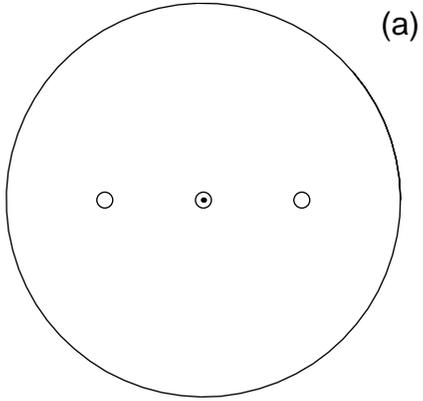
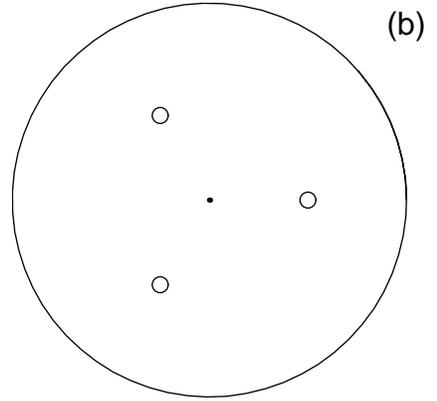
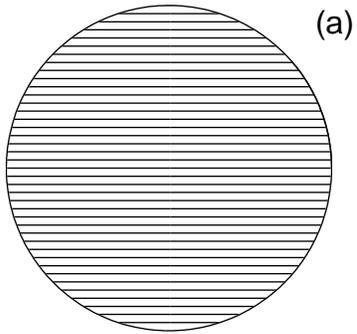
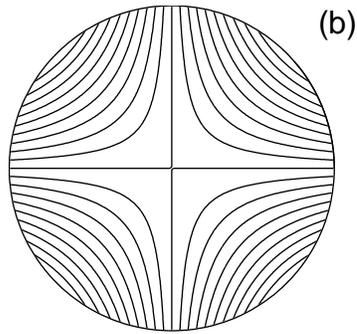
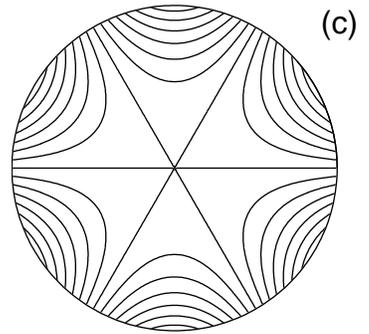
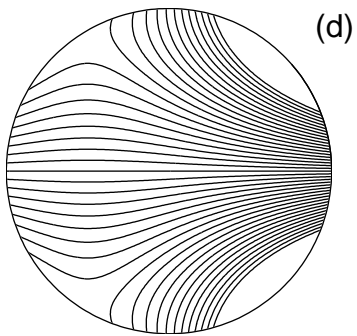
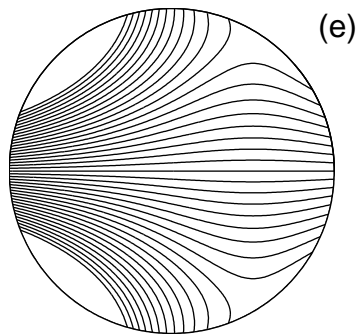
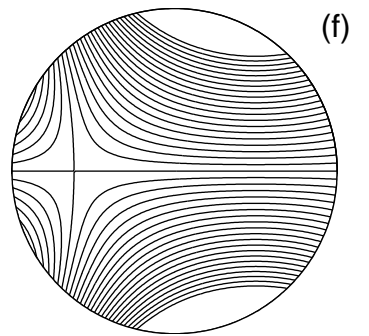
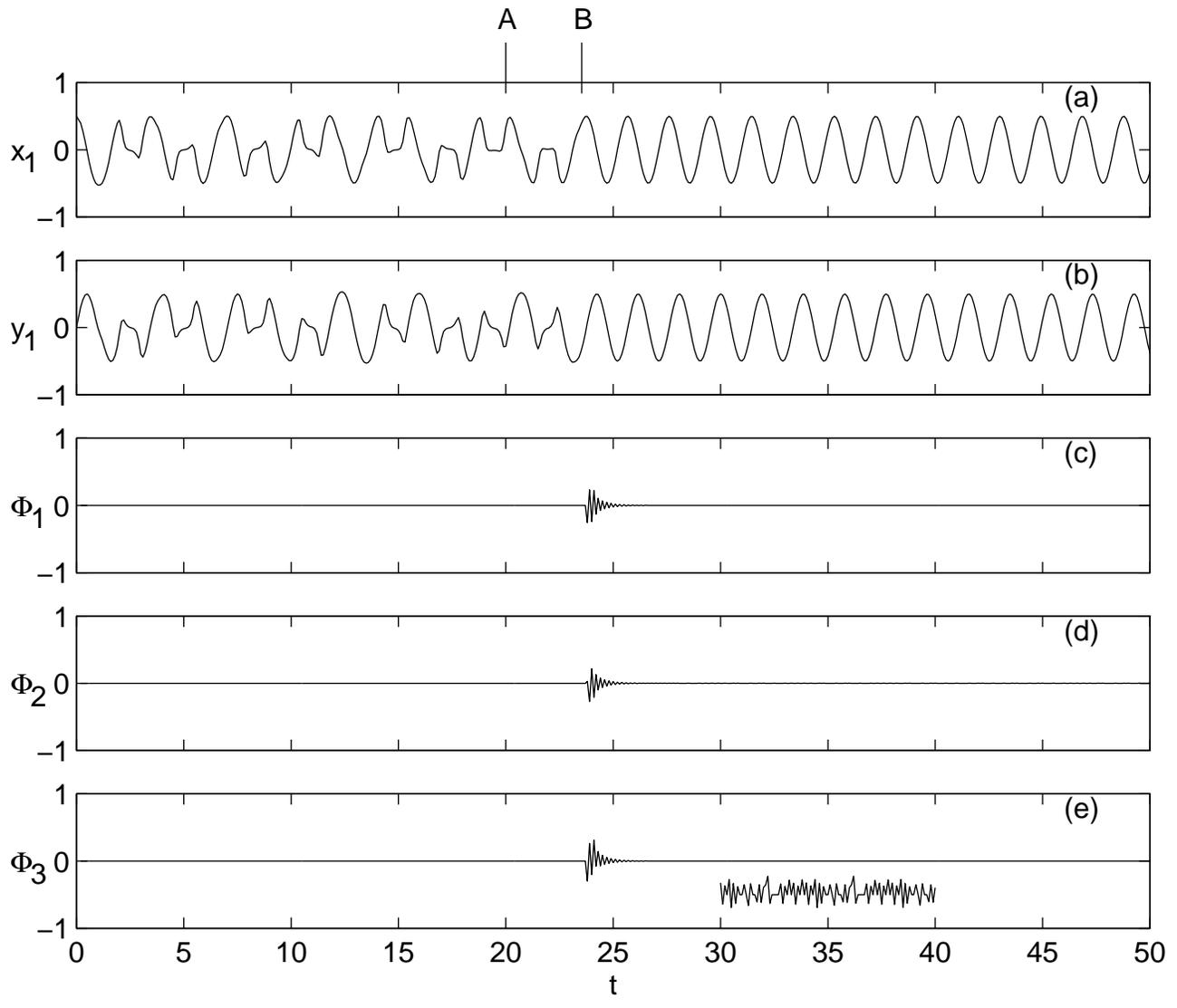
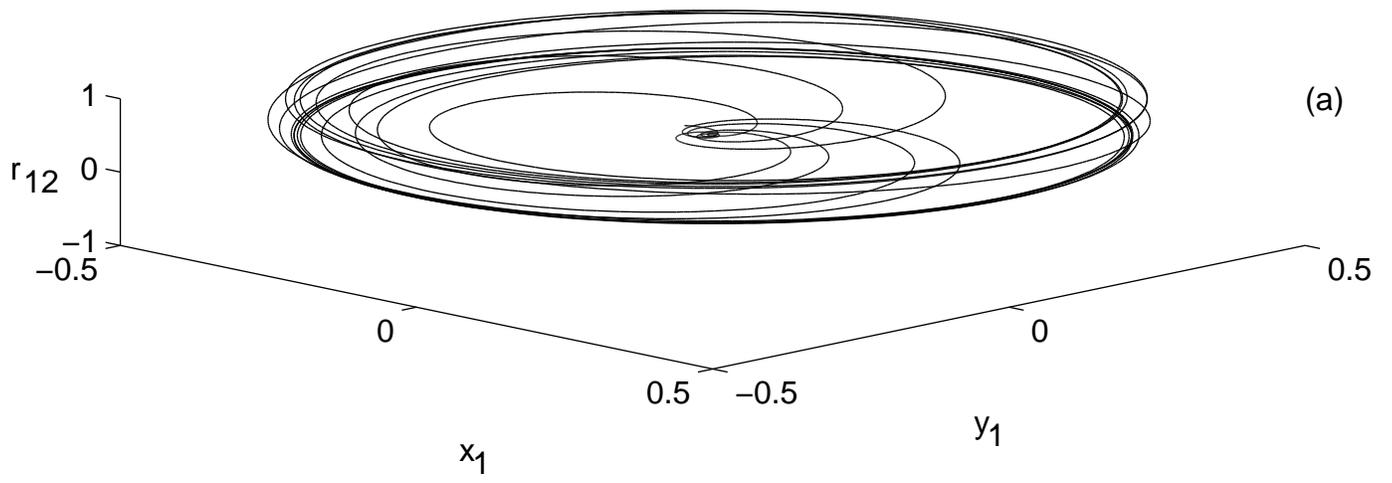
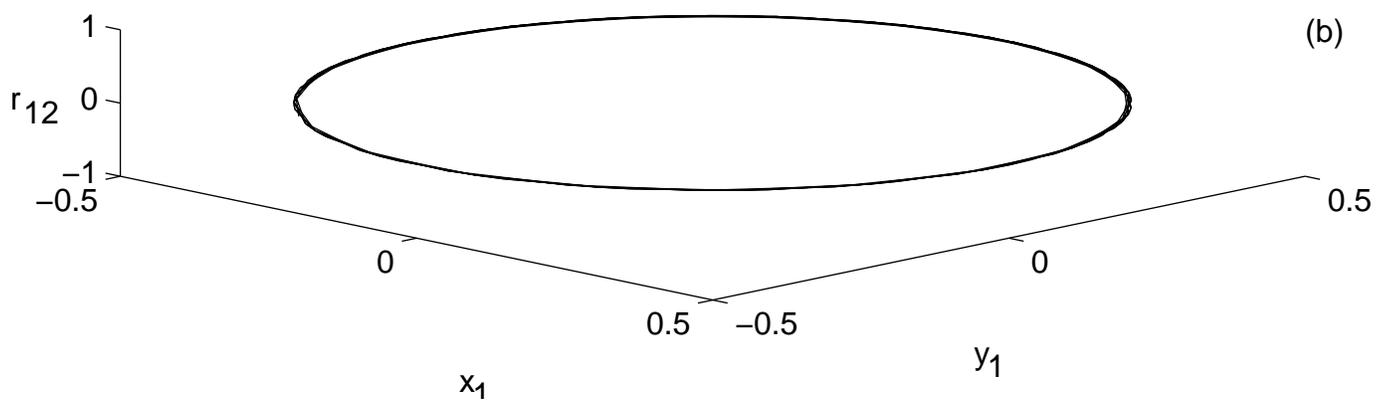
\end{document}